\documentclass{iaus}
\usepackage{graphicx}

\title[SAURON Absorption Lines of Bulges] 
{The Nature of Galactic Bulges from SAURON Absorption Line Strength Maps}

\author[Peletier, Falc\'on-Barroso and the SAURON team]   
{Reynier F. Peletier$^1$,
Jes\'us Falc\'on-Barroso$^2$, Katia Ganda$^1$, \break 
Roland Bacon$^{3}$, Michele Cappellari$^{4}$, Roger L. Davies$^{4}$, \break 
P. Tim de Zeeuw$^{5}$, Eric Emsellem$^{3}$, Davor Krajnovi\'c$^{4}$, \break
Harald Kuntschner$^{6}$, Richard M. McDermid$^{5}$, 
Marc Sarzi$^{7}$, \and Glenn van de Ven$^{8}$}

\affiliation{$^1$Kapteyn Astronomical Institute, University of Groningen,
NL-9700 AV Groningen \break email:
peletier@astro.rug.nl \\[\affilskip]
$^2$European Space Agency / ESTEC, Keplerlaan 1, NL-2200 AG Noordwijk, \break
$^3$Observatoire de Lyon, 9 av. Charles Andr\'e,
F-69230 Saint-Genis Laval,\\
$^4$Sub-Department of Astrophysics, University of Oxford, Oxford OX1 3RH, UK,
\break
$^5$Sterrewacht Leiden, University of Leiden, NL-2333~CA Leiden, \break
$^6$ST-ECF, European Southern Observatory, D-85748 Garching bei M\"unchen,
\break
$^7$Centre for Astrophysics Research, University of Hertfordshire, Hatfield, UK,
\break
$^8$Institute for Advanced Study, Einstein Drive, Princeton, NJ 08540, USA
} 

\pubyear{2007}
\volume{241}  
\pagerange{1-4}
\jname{Proceedings Stellar Populations as Building Blocks of Galaxies}
\editors{A. Vazdekis \& R.F. Peletier, eds.}
\begin{document}

\maketitle

\begin{abstract}

We discuss SAURON absorption line strength maps of a sample of  24 early-type
spirals, mostly Sa. From the Lick indices H$\beta$, Mg\,$b$ and Fe 5015 we
derive SSP-ages and metallicities. By comparing the scaling  relations  of
Mg\,$b$ and H$\beta$ and central velocity dispersion with the same relation for
the edge-on sample of Falc\'on-Barroso et al. (2002) we derive a picture in
which the central regions of Sa galaxies contain at least 2 components:  one
(or more) thin, disc-like component, often containing recent star formation,
and another, elliptical-like  component, consisting of old stars and rotating
more slowly, dominating the light above the plane. If one defines a bulge to be
the component responsible for the light in excess of the outer exponential
disc, then many Sa-bulges are dominated by a thin, disc-like component
containing recent star formation.
\end{abstract}

\firstsection 
\section{Introduction}

The measurement of absorption line strengths in combination with stellar
population models has been used for many years to probe the ages and 
metallicities from integrated  stellar populations of galaxies. Although there
exist many studies of elliptical galaxies and S0's, absorption line studies of
spiral galaxies are lagging behind. They are, however, important to understand
the origin of galactic bulges. Are bulges old, elliptical-like objects in the middle
of a large, spiral disk, which formed first? Or are they mass and light
concentrations that formed from internal processes in the disk? Recent reviews
about this topic are given by Kormendy \& Kennicutt (2004) and Athanassoula
(2005). Stellar population studies of early-type spirals (Jablonka et al. 1996,
Proctor \& Sansom 2002) see very little difference between the central stellar
populations of spirals and S0's. Both studies, however, contain few fainter
galaxies with a central velocity dispersion smaller than 120 km/s. Those
objects show stellar populations with a variety of properties (Moorthy \&
Holtzman 2006). At present it looks as if both types of bulges (elliptical-like
and disk-like) exist. It is not clear, however, why this is, and how the
bulge-type is related to its stellar populations.

In this work we present high quality, two-dimensional absorption line maps of a
sample of 24 early-type spirals. Here some highlights of the work are given.
More details are found in Peletier et al. (2007).

\section{SAURON absorption line strength maps of Sa galaxies}

We have obtained Integral Field Spectroscopy in a field of 33$''$ $\times$
41$''$, with a spatial sampling of 0.94$''$ $\times$ 0.94$''$ using
SAURON at the WHT in La Palma. The observations are part of the SAURON survey,
described in de Zeeuw et al. (2002). The sample consists of 24 early-type 
spiral galaxies, for which the kinematics of gas and stars have been presented
in Falc\'on-Barroso et al. (2006). The spectra, which have a wavelength range from 4790
to 5300 \AA,  were fitted with the stellar population models of Vazdekis (1999), 
allowing us to separate the emission lines
from the absorption line spectrum (for details about this procedure see Sarzi
et al. 2006 and Falc\'on-Barroso et al. 2006). From the cleaned spectra we 
obtained the line indices H$\beta$, Mg\,$b$ and Fe 5015. 

In the way we described in Kuntschner et al. (2006) we determined ages,
metallicities and abundance ratios at every position, assuming that the stellar
populations there could be represented by a single-age, single metallicity
stellar population. In practise, we determined the SSP for which the line
strengths Fe 5015 , H$\beta$ and Mg\,$b$ fitted best in the $\chi^2$ sense
(Fig.~1). 
Although we know that it is a great
over-simplification to represent the stellar populations (even locally) of a
galaxy by a SSP, in some, especially elliptical
galaxies this approach gives a good first-order approximation.

\begin{figure}
 \includegraphics[scale=0.83]{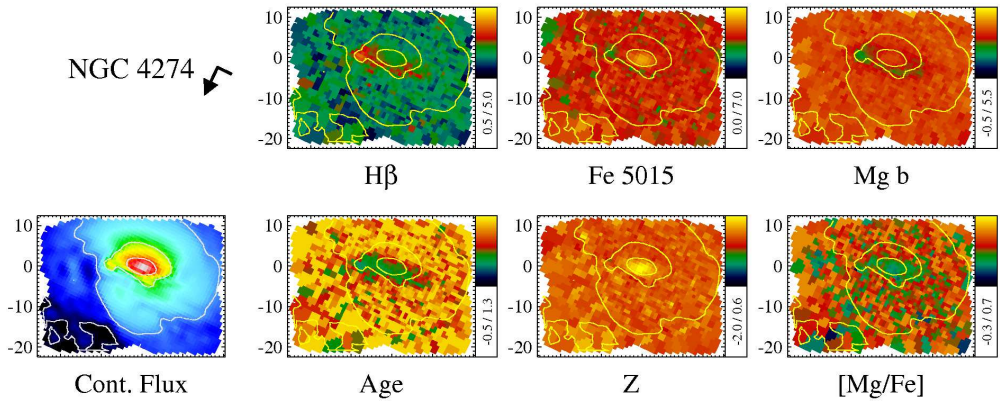}
 \caption{Absorption line strength maps of one of the galaxies.
Shown are (from left to right): 
line indices H$\beta$ ,Fe 5015  and Mg\,$b$.
Second row: Reconstructed intensity, logarithmic Age, Metallicity 
(log Z/Z$_\odot$) and [$\alpha$/Fe]. The reconstructed intensity is overlaid in 
white contours on the maps. }\label{fig:peletier1}
\end{figure}

\section{Relations between indices and velocity
dispersion}


Early-type galaxies show a tight Mg$_2$ -- $\sigma$ relation (Terlevich et al.
1981). Deviations from the relations correlate well with parameters indicating
the presence of young stellar populations. In
Falc\'on-Barroso et al. (2002) we used the relation to show that the stellar
populations in a sample of inclined early-type spirals are generally old. In
Fig. 2 we show the central Mg\,$b$ and H$\beta$ indices of  our sample as a
function of the central velocity dispersion $\sigma_{\rm cen}$.  In the figure
are shown the galaxies of this sample, together with the ellipticals and
lenticulars of Kuntschner et al. (2006) (at r$_e$/8), and a number of
literature samples of  early-type spirals  (see caption).
The black line is a best fit to the ellipticals and S0 galaxies in the Coma cluster
of J\o rgensen et al. (1996). The Mg\,$b$  - $\sigma$ relation of elliptical
galaxies and S0's acts as an upper envelope for the Sa galaxies. Although  some
Sa galaxy  centre measurements lie close to the relation, a significant fraction of the galaxies
falls below it. The same effect is seen for the H$\beta$ - $\sigma$ relation. 
Using the argumentation of Schweizer et al.,  the line of galaxies in
Coma would correspond to old stellar populations, while deviations would be caused by
younger stars. The fact that our Sa bulges  mostly lie below the Mg\,$b$ - $\sigma$
relation or above the  H$\beta$ - $\sigma$ relation would indicate that the centres of Sa
bulges generally are significantly younger than  early-type galaxies in the Coma
cluster.
This result appears to contradict the tight Mg$_2$ -- $\sigma$ relation for bulges
found by FB02 and also the relation by Jablonka et al.
(1996). It confirms, however, the results of Prugniel et al. (2001), 
who find several early-type spiral galaxies
lying considerably below the Mg$_2$ -- $\sigma$ relation. Notice that there are 
several S0 galaxies that are far away from the relation defined by elliptical
galaxies, in the same location  as the spirals with the lowest Mg\,$b$ values.


\begin{figure}
 \includegraphics[scale=0.9]{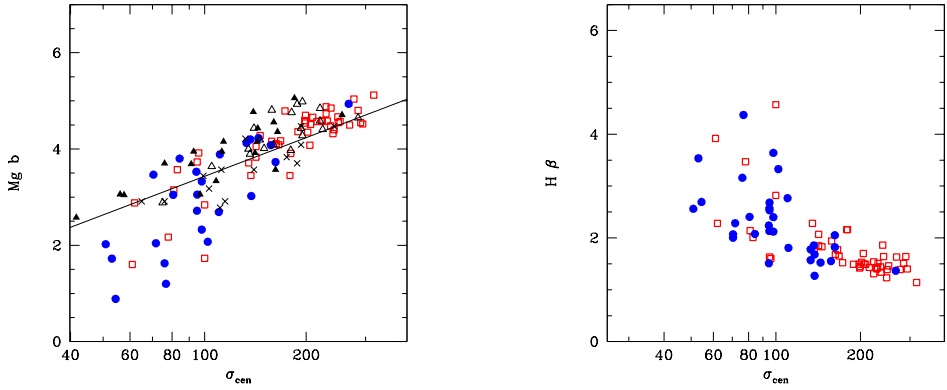}
  \caption{{\bf Left: }Central indices as a function of central velocity \
dispersion (in km/s). The open red symbols show the ellipticals and S0 galaxies
of Kuntschner et al. (2006) for an aperture of $r_e/8$. 
The black line is the
least-squares fit to the ellipticals and S0 galaxies in Coma of J\o rgensen et
al. (1996). As a comparison we also show a few literature samples in black:
the highly-inclined bulges of FB02 (filled triangles), the bulges of 
Bender et al. (1993, open triangles), and bulges of
Jablonka et al. (1996, crosses).}\label{fig:peletier2}
\end{figure}

\section{Star formation histories in the central regions of early-type spirals}

In the region of interest ($\sigma$ $<$ 120 km/s) the galaxies of  FB02 
generally have higher Mg\,$b$ than the galaxies of this sample. Why this
difference? The only important difference between the two samples is the
inclination distribution. If Sa galaxies would contain young stellar 
populations that would be situated in the plane, we would see them in the
SAURON sample. In FB02, however, where we looked at 5$''$ from the center on
the minor axis, we would not have seen them, if the young stellar populations
were limited to the very central regions. 
Since for these Sa galaxies the light
in the central regions is completely dominated by the bulge, 
it seems that bulges consist
of 2, maybe more components: one which is old, elliptical-like, and slowly 
rotating, and another more flattened, disk-like component containing often
young stellar populations. Support to this idea is given by the fact that the
SAURON maps show that the young stellar populations can often be found in
circumnuclear rings, features with a small vertical extent. Another supporting
argument is the fact that many of the galaxies have local central velocity
dispersion minima (Falc\'on-Barroso et al. 2006, see also Ganda et al. 2006,
Emsellem et al. 2001). These are most likely caused by central discs, some of
which contain young stellar populations.

According to the current literature there are several kinds of bulges. Bulges
that photometrically (r$^{1/4}$ surface brightness law) and kinematically
(still relatively high $\lambda_R$, Emsellem et al. 2007) resemble
elliptical galaxies are often called classical bulges (Kormendy \& Kennicutt
2004). A bulge consisting only of the fast-rotating component is called a
pseudo-bulge in this reference. Athanassoula
(2005) claims that there are three types of bulges: the classical  bulges,
which form by collapse or merging,  disc-like bulges, which result from the
inflow of (mainly) gas to the  centre-most parts, and subsequent star
formation, and boxy and peanut bulges, which are seen in  near-to-edge-on
galaxies and which are in fact just a part of the bar seen edge-on,  and
therefore not part of the bulge in the definition of this paper.
Here we add another piece of the puzzle. From the stellar population
distribution,  by comparing a sample uniformly distributed in inclination with
a sample biased towards  high inclination we infer that galactic bulges have
more than one physical component: generally they have a slowly-rotating,
elliptical-like component, and one or more fast-rotating components in the
plane of the galaxy.  This picture also nicely explains the fact that bulge
populations in general are very similar to those in the disc (e.g. Peletier \&
Balcells 1996).

\end{document}